\documentclass[12pt]{article}

\usepackage[hmargin=2.0cm, vmargin=2.3cm]{geometry}
\usepackage[numbers,super,sort&compress]{natbib}
\usepackage{hyperref}

\usepackage{setspace}

\usepackage{graphicx}
\usepackage{amsfonts}
\usepackage{amssymb}
\usepackage{amsmath}
\usepackage{multibib}
\newcites{supp}{Additional References for Methods}

\usepackage{graphicx}
\usepackage{epsfig}
\usepackage{epstopdf}
\usepackage{float}
\usepackage{url}

\graphicspath{{./}}

\begin{document}

\title{The solar dynamo begins near the surface\footnote{Accepted in Nature}}

\author{
Geoffrey M.~Vasil$^{1}$\footnote{\url{gvasil@ed.ac.uk}}, 
Daniel Lecoanet$^{2,3}$, 
Kyle Augustson$^{2,3}$, 
\\ 
Keaton J.~Burns$^{4,5}$, 
Jeffrey S.~Oishi$^{6}$, 
Benjamin P.~Brown$^{7}$, 
\\
Nicholas Brummell$^{8}$,
Keith Julien$^{9}$ \\ 
\\
$^{1}$School of Mathematics and  the Maxwell Institute for Mathematical Sciences, \\
University of Edinburgh, EH9 3FD, UK\\ \\ 
$^{2}$Department of Engineering Sciences  and Applied Mathematics, \\ Northwestern University, Evanston IL 60208, USA\\ \\ 
$^{3}$CIERA, Northwestern University, Evanston IL 60201, USA\\ \\ 
$^{4}$Department of Mathematics,  Massachusetts Institute of Technology,  \\ Cambridge MA 02139, USA\\ \\ 
$^{5}$Center for Computational Astrophysics,  Flatiron Institute, \\ New York NY 10010, USA\\ \\ 
$^{6}$Department of Physics \& Astronomy, Bates College, \\ Lewiston ME 04240, USA\\ \\ 
$^{7}$Department of Astrophysical and Planetary Sciences, \\ 
University of Colorado Boulder, Boulder CO 80309, USA\\ \\ 
$^{8}$Department of Applied Mathematics, Jack Baskin School of Engineering,\\ University of California Santa Cruz, Santa Cruz, CA 95064, USA\\ \\ 
$^{9}$Department of Applied Mathematics, \\ 
University of Colorado Boulder, Boulder CO 80309, USA\\ \\
}

\date{\today}

\maketitle

\doublespacing

{\bf
The Sun's magnetic dynamo cycle features a distinct pattern: a propagating region of sunspot emergence appears around $\boldsymbol{30^{\circ}}$ latitude and vanishes near the equator every 11 years.
Moreover, longitudinal flows called ``torsional oscillations" closely shadow sunspot migration, undoubtedly sharing a common cause\cite{Snodgrass1985}. 
Contrary to theories suggesting deep origins for these phenomena, helioseismology pinpoints low-latitude torsional oscillations to the Sun's outer 5-10\%, the ``Near-Surface Shear Layer"\cite{Vorontsov2002, Hathaway2022}. 
Within this zone, inwardly increasing differential rotation coupled with a poloidal magnetic field strongly implicates the Magneto-Rotational Instability\cite{Chandrasekhar1960,Balbus1991} prominent in accretion-disk theory and observed in laboratory experiments\cite{Wang2022}.
Together, these two facts prompt the general question: Is it possible that the solar dynamo is a near-surface instability? 
Here, we report strong affirmative evidence in stark contrast to traditional paradigms\cite{Parker1993} focusing on the deeper tachocline.
Simple analytic estimates show that the near-surface magneto-rotational instability better explains the spatiotemporal scales of the torsional oscillations and inferred subsurface magnetic field amplitudes\cite{Baldner2009}. 
State-of-the-art numerical simulations corroborate these estimates and, strikingly, reproduce hemispherical magnetic current helicity laws\cite{Pevtsov1995}. 
The dynamo resulting from a well-understood near-surface phenomenon improves prospects for accurate predictions of full magnetic cycles and space weather, impacting Earth's electromagnetic infrastructure.
}

\bigskip 

Key observations any model must regard include:
\textit{the solar butterfly diagram}, a decadal migration pattern of sunspot emergence\cite{Maunder1922,Hathaway2022} with strong latitude dependence; 
\textit{the torsional oscillations} constituting local rotation variations corresponding with magnetic activity\cite{Snodgrass1985,Vorontsov2002,Hathaway2022};
\textit{the poloidal field}, an $\approx 1\,\text{G}$ photospheric field  with a 1/4-cycle phase lag relative to sunspots\cite{Babcock1961}, and $\approx 100\, \text{G}$ subsurface amplitudes\cite{Baldner2009};
\textit{the hemispherical helicity sign rule} comprising an empirically observed negative current helicity in the northern hemispheres and positive current helicity in the south\cite{Pevtsov1995}; 
\textit{the tachocline} at the base of the convection zone, the traditionally proposed seat of the solar dynamo; and 
\textit{the near-surface-shear layer (NSSL)} within the Sun's outer 5-10\% containing strong inwardly increasing angular velocity fostering the Magneto-Rotational Instability (MRI).

Despite progress, prevailing theories have distinct limitations.
\textit{Interface dynamos} (proposed within the tachocline\cite{Parker1993}) preferentially generate high-latitude fields\cite{Karak2017}, and would produce severe shear disruptions\cite{Vasil2009} which are not observed\cite{Howe2020}. 
\textit{Mean-field dynamos} offer qualitative insights but suffer from the absence of first principles\cite{Cattaneo2006}, and are contradicted by observed meridional circulations\cite{Chen2017}.
\textit{Global convection-zone models} often misalign with critical solar observations, require conditions diverging from solar reality\cite{Nelson2011,Kpyl2014,Hotta2021}, and fail to provide a theoretical dynamical understanding. 

Borrowing from well-established ideas in accretion-disk physics\cite{Chandrasekhar1960,Balbus1991}, we propose an alternative hypothesis that produces clear predictions and quantitatively matches key observations.

For electrically conducting plasma like the Sun, the local axisymmetric linear instability criterion for the MRI is\cite{Chandrasekhar1960, Balbus1991}:
\begin{equation}
2\Omega S  \ > \  \omega_{A}^{2}  \label{MRI-condition},
\end{equation}
where the local Alfv\'{e}n frequency and shear are
\begin{equation}
\omega_{A} = \frac{B_{0}\, k_{r}}{\sqrt{4\pi \rho_{0}}} \qquad \text{and} \qquad S = -r \, \frac{d\Omega}{dr}.
\label{freq-definition}
\end{equation}
The system control parameters are the background poloidal magnetic field strength ($B_{0}$ in cgs units), the atmospheric density ($\rho_{0}$), the smallest non-trivial radial wavenumber that will fit in the domain ($k_{r} \approx \pi/H_{r}$, where $H_{r}$ is a relevant layer depth or density-scale height), bulk rotation rate ($\Omega$), and the differential rotation, or shear ($S > 0$ in the NSSL). An adiabatic density stratification holds to a good approximation within the solar convection zone, eliminating buoyancy modifications to the stability condition (\ref{MRI-condition}).  

The MRI is essential for generating turbulent angular momentum transport in magnetized astrophysical disks\cite{Balbus1991}. Previous work\cite{Brandenburg2005} postulated the NSSL as the possible seat of the global dynamo without invoking the MRI. Another kinematic-dynamo study\cite{Dikpati2002} dismissed the relevance of NSSL for kinematic dynamos but did not allow for full magnetohydrodynamic (MHD) instabilities (such as the MRI). Modern breakthroughs in our understanding of large-scale MRI physics\cite{Vasil2015,Oishi2020} have not been applied in a solar context, and local MRI studies of the Sun\cite{Kagan2014} have only considered small scales. To our knowledge, no work has yet considered the large-scale MRI dynamics relevant to the observed features of the global dynamo. We, therefore, describe here a potential MRI-driven solar dynamo cycle.

The start of the solar cycle is the period surrounding the sunspot minimum when there is no significant toroidal field above the equator and a maximal poloidal field below the photosphere.
This configuration is unstable to the axisymmetric MRI, which generates a dynamically active toroidal field in the outer convection zone.
The observed torsional oscillations are the longitudinal flow perturbations arising from the MRI. The relative energetics are consistent with nonlinear dynamo estimates (see Methods). As the cycle progresses, the toroidal field can undergo several possible MHD instabilities contributing to poloidal-field regeneration, e.g., the helical MRI, non-axis-symmetric MRI, the ``clamshell'' instability, and several more, including a surface Babcock-Leighton process. 
We hypothesise that the axisymmetric subsurface field and torsional oscillations constitute a nonlinear MRI travelling wave.
The instability saturates via radial transport of (globally conserved) mean magnetic flux ($B_{0}$) and angular momentum ($\Omega, S$), which neutralise the instability criterion eq.~(\ref{MRI-condition}) (see Methods). 

Empirical timescales of the torsional oscillations imply an approximate growth rate, $\gamma$, for the MRI and, hence, a relevant poloidal field strength. 
To a good approximation, $S \approx  \Omega \approx  2\pi/\text{month}$ in the NSSL (see figs.~1(a)-(c)).
The early-phase torsional oscillations change on a timescale of 2--12 months,
implying a growth rate of $\gamma/\Omega \approx  \text{ 0.01--0.1}$ (see Methods).
A modest growth rate and the regularity of the solar cycle over long intervals together suggest that the global dynamics operate in a mildly nonlinear regime.
Altogether, we predict roughly $\omega_{A} \approx  S \approx  \Omega$.

The torsional oscillation pattern shows an early-phase mode-like structure with an approximately 4:1 horizontal aspect ratio occupying a depth of roughly $r/R_{\odot} \approx  5\%$, or $k_{r} \approx 70\, R_{\odot}^{-1}$; see fig.~1(d).
Using eq.~(\ref{freq-definition}), the approximate background Alfv\'{e}n speed $v_{A} \approx  \text{200--2000}\, \text{cm/s} \approx  \text{0.1--1.0}\, R_{\odot}/\text{year}$.

Alfv\'{e}n-speed estimates required for MRI dynamics are consistent with observed internal magnetic field strengths.
Measurements suggest $\approx\text{100--200}\,\text{G}$ internal poloidal field\cite{Baldner2009}, agreeing with the above estimates using NSSL densities $\rho_{0} \approx  3\!\cdot\! 10^{-2}\,$--$\,3\!\cdot\! 10^{-4}\,\text{g}/\text{cm}^{3}$.
The same studies found roughly similar ($\approx\text{300--1000}\,\text{G}$) internal toroidal field strength confined within the NSSL.
Given solar-like input parameters, a detailed calculation finds the MRI should operate with latitudinal field strengths up to $\approx1000\,\text{G}$ (see Methods). 

Background shear modification dominates the MRI saturation mechanism (see Methods), roughly 
\begin{equation}\label{saturation}
    |\Omega'|^{2} \approx \frac{H_{r}^{2}}{R_{\odot}^{2}}\frac{(2 \Omega S - \omega_{A}^{2}) \left(S^2+\omega_{A} ^2\right)^2}{2\Omega (S+2 \Omega ) \left(S^2+(2 \Omega) ^2 + 2
   \omega_{A} ^2\right)},
\end{equation}
where $\Omega'$ represents the dynamic changes in differential rotation. 
For $S\approx \Omega \approx \omega_A$, $|\Omega'| \approx 7\, {\rm nHz}$, roughly consistent with the observed torsional oscillation amplitude; see fig.~1(d).

We compute a suite of growing global perturbations using Dedalus\cite{Burns2020} to model the initial phase of the solar cycle with quasi-realistic solar input parameters (see Methods). Fig.~2 shows representative solutions.

We find two distinct cases: (i) a ``fast branch'' with direct growth rates, $\gamma$, comparable to \textit{a priori} estimates; and (ii) a ``slow branch'' with longer but relevant growth times and oscillation periods.
The eigenmodes are confined to the NSSL, reaching from the surface to $r/R_{\odot} \approx  0.90\text{--}0.95$, where the background shear becomes MRI stable.

For case (i), $\gamma/\Omega_{0} \approx 6\! \cdot\! 10^{-2}$ (given $\Omega_{0}=466\,$nHz) with corresponding $e$-folding time, $t_{e} \approx 60\,$days and no discernible oscillation frequency. The pattern comprises roughly one wave period between the equator and $\approx 20^{\circ}$ latitude, similar to the rotation perturbations seen in the torsional oscillations.

For case (ii), $\gamma/\Omega_{0} \approx 6\! \cdot\! 10^{-3}$ with $t_{e} \approx 600\,$days and oscillation frequency $\omega/\Omega_{0} \approx 7\! \cdot\! 10^{-3}$, corresponding to a period $P\approx 5\,$years. The pattern comprises roughly one wave period between the equator and $\approx 20^{\circ}\text{--}30^{\circ}$ latitude. 

In addition to cases (i) and (ii), we find 34 additional purely growing ``fast-branch'' modes, two additional oscillatory modes and one intermediate exceptional mode (see Extended Data figs.~1--3). 

Using the full numerical MHD eigenstates, we compute a systematic estimate for the saturation amplitude using quasi-linear theory (see Methods): $|\Omega'|\approx 6\, {\rm nHz}$ for case (i) and $|\Omega'|\approx 3\, {\rm nHz}$ for the slow-branch case (ii); both comparable to the observed torsional oscillation amplitude and the simple analytical estimates from eq.~(\ref{saturation}). The true saturated state would comprise an interacting superposition of the full spectrum of modes.

Notably, the slow-branch current helicity, $\mathcal{H} \propto \boldsymbol{b} \boldsymbol{\cdot} \boldsymbol{\nabla\times} \boldsymbol{b}$, follows the hemispherical sign rule\cite{Pevtsov1995}, with $\mathcal{H} < 0$ in the north and $\mathcal{H} > 0$ in the south.
The slow-branch modes appear rotationally constrained, consistent with their low Rossby number\cite{Vasil2021}, perhaps providing a pathway for understanding the helicity sign rule. 

New helioseismic data analyses could test our predictions.
The MRI would not operate if the poloidal field is too strong, nor would it explain the torsional oscillations if it is too weak.
We predict correlations between the flow perturbations and magnetic fields, which time-resolved measurements could test, constraining joint helioseismic inversions of flows and magnetic fields.

An MRI-driven dynamo may also explain the formation and cessation of occasional grand minima\cite{Eddy1976} (e.g., Maunder).
An ``essentially nonlinear" MRI dynamo does not start from an infinitesimal seed field upon each new cycle (see Methods). 
Rather, a moderate poloidal field exists at the solar minimum, and the MRI processes it into a toroidal configuration.  
If the self-sustaining poloidal-to-toroidal regeneration sometimes happens imperfectly, then subsequent solar cycles could partially fizzle, leading to weak subsurface fields and few sunspots. Eventually, noise could push the system back onto its normal cyclic behaviour, as in the El Nino Southern Oscillation\cite{Suarez1988}.

Finally, our simulations intentionally contain reduced physics to isolate the MRI as a critical agent in the dynamo process, filtering out large-scale baroclinic effects, small-scale convection and nonlinear dynamo feedback.
Modelling strong turbulent processes is arduous: turbulence can simultaneously act as dissipation, drive large-scale flows like the NSSL, produce mean electromotive forces, and excite collective instabilities. 
While sufficiently strong turbulent dissipation could eventually erase all large-scale dynamics, the mere presence of the solar torsional oscillations implies much can persist within the roiling background.

\pagebreak

\begin{figure}[H]
\begin{center}
\begin{tabular}{c} \\
\includegraphics[width=16cm]{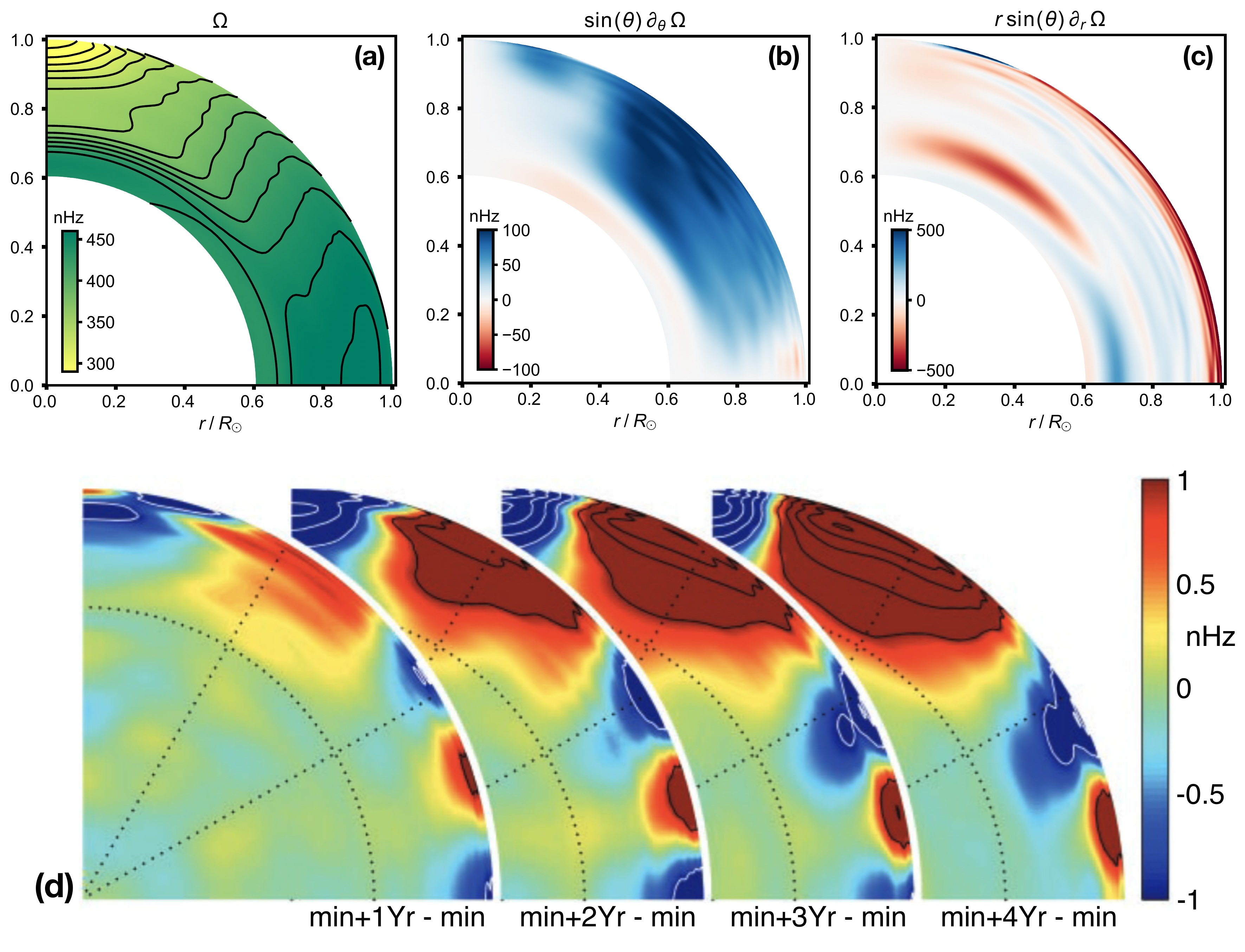}
\end{tabular}
\end{center}
\vspace{0.0cm}
\singlespace
\caption{Measured internal solar rotation profiles. \textbf{a}, Heliosesimic differential rotation profile, $\Omega(\theta,r)$, using publicly available data from [T. P. Larson and J. Schou. Global-mode analysis of full-disk data from the Michelson Doppler imager and the helioseismic and magnetic imager. \textit{Solar Physics}, 293(2), 2018]\cite{Larson2018}. \textbf{b, c} The respective latitudinal and radial shear gradients $r \sin(\theta) \boldsymbol{\nabla} \Omega(\theta,r)$; computed via a non-uniform 4th-order centered finite-difference scheme. The latitudinal mean of tachocline shear is $\approx \! 200\,$nHz and peak amplitudes are below $\approx\! 350\,$nHz. Conversely, the near-surface shear averages $\approx 400$-$600\,$nHz (with rapid variation in depth) and peak values over $1200\,$nHz.
Bottom row: \textbf{d}, Helioseismic measurements of solar torsional oscillations. The red shows positive residual rotation rates and blue shows negative residual rotation rates after removing the 1996 annual mean of $\Omega(r,\theta)$. Each slice shows the rotational perturbations 1, 2, 3 and 4 years after the approximate solar minimum.  
The notation ``min+1Yr - min'' means taking the profile at 1 year past solar minimum and subtracting the profile at solar minimum. 
The colour table saturates at $1\,$nHz, corresponding to about $\approx 400\,$cm/s surface flow amplitude. Further contour lines show 1 nHz increments within the saturated regions. 
Diagram in \textbf{d} reprinted from figure 2 of [S.~V.~Vorontsov, J.~Christensen-Dalsgaard, J.~Schou, V.~N.~Strakhov, and M.~J.~Thompson.
Helioseismic measurement of solar torsional oscillations. \textit{Science}, 296(5565):101–103, April
2002.]\cite{Vorontsov2002} with permission from AAAS.}
\end{figure}

\pagebreak

\begin{figure}[H]
\begin{center}
\begin{tabular}{c} \\
\includegraphics[width=13cm]{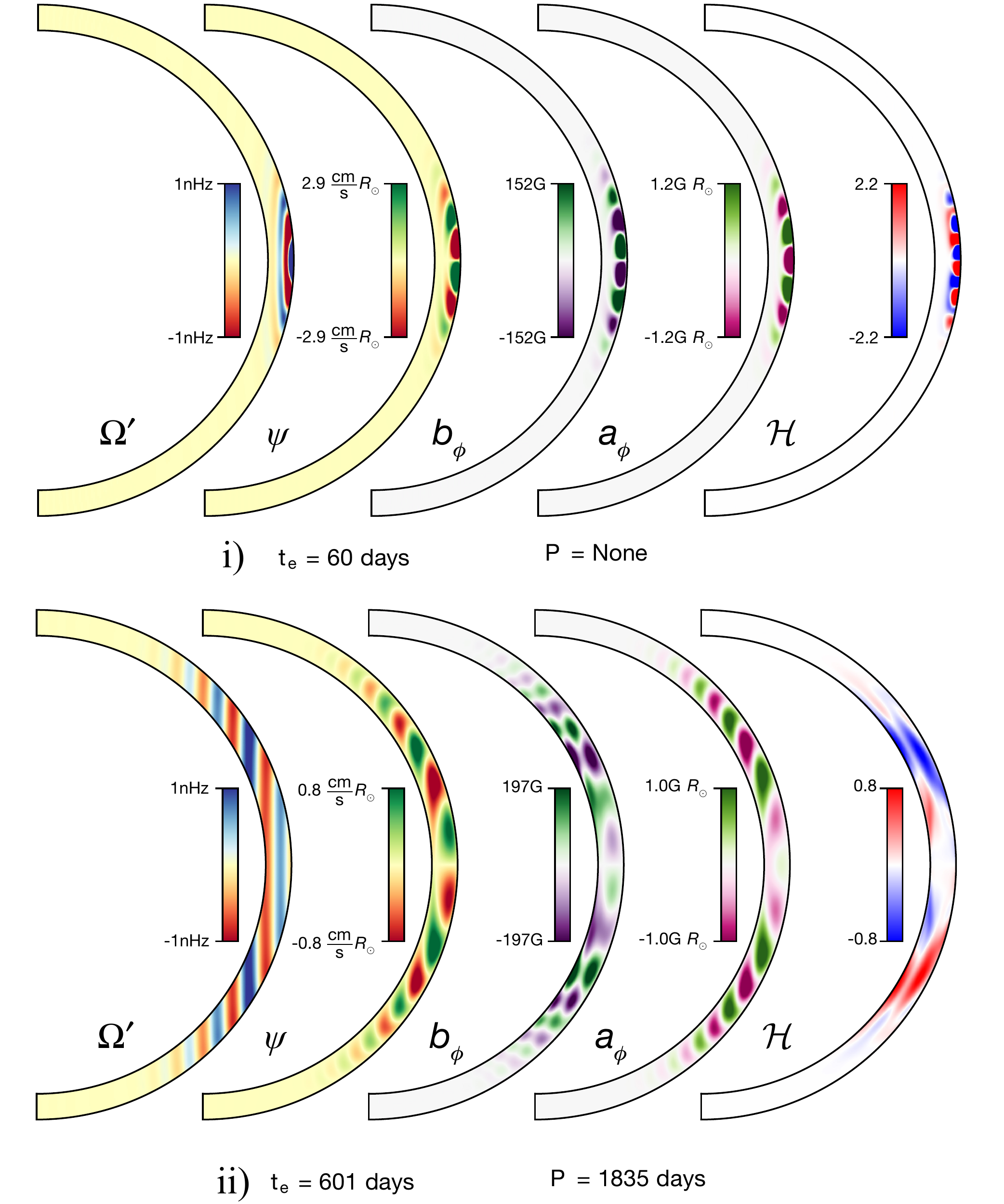}
\end{tabular}
\end{center}
\vspace{0.0cm}
\singlespace 
\caption{Two meridional ($r,\theta$) MRI eigenmode profiles. In both cases (i) \& (ii) from left to right: longitudinal angular velocity perturbation, $\Omega'(r,\theta) = u_{\phi}(r,\theta)/(r\sin(\theta))$; momentum-density streamfunction ($\phi$-directed component; see Methods), $\psi(r,\theta)$; longitudinal magnetic field, $b_{\phi}(r,\theta)$; magnetic scalar potential, $a_{\phi}(r,\theta)$; current helicity correlation, $\mathcal{H}(r,\theta)$. The timescales $t_{e}, P$ represent the instability $e$-folding time and oscillation period, respectively. Top row: Case (i) shows a typical directly growing ``fast-branch'' mode with no oscillation and growth rates $\gamma \approx 0.06 \, \Omega_{0}$. 
The bottom row: Case (ii) shows a typical large-scale ``slow-branch'' mode with a roughly five-year period. In each case, we fix the overall amplitude to $1\,{\rm nHz}$ for the rotational perturbations, with all other quantities taking their corresponding relative values.}
\end{figure}

\section*{Methods}

\subsection*{Numerical calculations}

We solve for the eigenstates of the linearised anelastic MHD equations\citesupp{Brown2012,Vasil2013} in spherical polar coordinates, $(r,\theta,\phi) = (\textit{radius},\textit{colatitude},\textit{longitude})$. 
Using $R_{\odot} = 6.96 \cdot \! 10^{10}\, \text{cm}$ for the solar radius,
we simulate radii between $r_{0} \le r \le r_{1}$ where $r_{0}/R_{\odot} = 0.89$ and  $r_{1}/R_{\odot} = 0.99$. We place the top of the domain at 99\% because several complicated processes quickly increase in importance between this region and the photosphere (e.g., partial ionisation, radiative transport, and much stronger convection effects). We use the anelastic MHD equations in an adiabatic background to capture the effects of density stratification on the background Alfv\'{e}n velocities (density varies by roughly a factor of 100 across the NSSL, causing a roughly factor of 10 change in Alfv\'{e}n speed) and asymmetries in velocity structures introduced by the density stratification via $\boldsymbol{\nabla \cdot}(\rho \boldsymbol{u})$.  A key aspect of the anelastic approximation is that all entropy perturbations must be small, which is reasonable in the NSSL below $0.99R_{\odot}$. We do not use the fully compressible equations, as these linear instability modes do not have acoustic components. Future MRI studies incorporating buoyancy effects (e.g., the deep MRI branches at high latitudes) should employ a fully compressible (but low Mach number) model\cite{Anders2023}. 

\textit{Input background parameters:} We include density stratification using a low-order polynomial approximation to the Model-S profile\citesupp{ChristensenDalsgaard1996}. In $\text{g}/\text{cm}^{3}$ units, with $\mathrm{z} =(r-r_{0})/(r_{1}-r_{0})$,
\begin{eqnarray}
\rho_{0} &=& \alpha_{0} - \alpha_{1}  \, \mathrm{z}  + \alpha_{2} \, \mathrm{z}^{2} - \alpha_{3} \, \mathrm{z}^{3} + \alpha_{4} \, \mathrm{z}^{4} \\
\alpha_{0} &=& 0.031256 \\
\alpha_{1} &=& 0.053193 \\
\alpha_{2} &=& 0.033703 \\
\alpha_{3} &=& 0.023766 \\
\alpha_{4} &=& 0.012326,
\end{eqnarray}
which fits the Model-S data to better than 1\% within the computational domain. The density at $\mathrm{z}=1$ is $\rho_{0} = 0.000326$ versus $0.031256$ at $\mathrm{z}=0$.

The density profile is close to an adiabatic polytrope with $r^{-2}$ gravity and $5/3$ adiabatic index.
An adiabatic background implies that buoyancy perturbations diffuse independently of the MHD and decouple from the system. 

We use a low-degree polynomial fit to the observed NSSL differential rotation profile.
For $\mu = \cos(\theta)$,
\begin{eqnarray}
\boldsymbol{u}_{0} &=&\Omega(r,\theta)\, r \sin(\theta)\,  \boldsymbol{e}_{\phi} \label{eq: omega_fit start} \\
\Omega(r,\theta) &=&   \Omega_{0}\,\mathrm{Z}(\mathrm{z})\, \Theta(\mu),
\end{eqnarray}
where $\Omega_{0} = 466\,\text{nHz} \approx 2.92 \! \cdot
\! 10^{-6}\,\text{s}^{-1}$ and 
\begin{eqnarray}
\mathrm{Z}(\mathrm{z}) &=& 1 + 0.02\, \mathrm{z} - 0.01\, \mathrm{z}^{2} - 0.03\, \mathrm{z}^{3} \\
\Theta(\mu) & = & 1 - 0.145 \,\mu^{2} - 0.148 \,\mu^{4}.  \label{eq: omega_fit end}
\end{eqnarray}
We use the angular fit from \citetsupp{Howe2009}. 
The radial approximation results from fitting the equatorial profile from \citet{Larson2018} shown in fig.~1(a). Below $60^{\circ}$ latitude, the low-degree approximation agrees with the full empirical profile to within 1.25\%. The high-latitude differential rotation profile is less constrained because of observational uncertainties. 

We define the background magnetic field in terms of a vector potential,
\begin{eqnarray}
\boldsymbol{B}_{0} &=& \boldsymbol{\nabla\times} \boldsymbol{A}_{0}  \\ 
\boldsymbol{A}_{0} &=&  \frac{\mathcal{B}(r)}{2}\, r  \sin(\theta )\, \boldsymbol{e}_{\phi},
\end{eqnarray}
where 
\begin{eqnarray}
\mathcal{B}(r) \ = \  B_{0}\,\left( ( r/r_{1})^{-3} -  (r/r_{1})^{2}\right),
\end{eqnarray}
and $B_{0} = 90\, \text{G}$.
The $r^{-3}$ term represents a global dipole. The $r^{2}$ term represents a field with a similar structure but containing electric current,
\begin{eqnarray}
\boldsymbol{J}_{0} = \frac{\boldsymbol{\nabla\times} \boldsymbol{B}_{0}}{4\pi} =  \frac{5B_{0}}{ 4\pi \,r_{1}^{2}} \, r  \sin(\theta) \, \boldsymbol{e}_{\phi}.
\end{eqnarray}
The background field is in MHD force balance, 
\begin{eqnarray}
\boldsymbol{J}_{0} \times \boldsymbol{B}_{0} \ = \ \boldsymbol{\nabla}( \boldsymbol{A}_{0} \boldsymbol{\cdot} \boldsymbol{J}_{0} ).
\end{eqnarray}
The MHD force balance generates magnetic pressure, which inevitably produces entropy, $s'$, and enthalpy, $h'$, perturbations via,
\begin{eqnarray}
\frac{\boldsymbol{\nabla} (\boldsymbol{A}_{0} \boldsymbol{\cdot} \boldsymbol{J}_{0} ) }{\rho_{0}}  + T_{0} \boldsymbol{\nabla} s' = \boldsymbol{\nabla} h',
\end{eqnarray}
where
\begin{eqnarray}
s' =  \frac{1}{\Gamma_{3}-1} \frac{\boldsymbol{A}_{0} \boldsymbol{\cdot} \boldsymbol{J}_{0}}{T_{0}\, \rho_{0} }, \qquad h' =  \frac{\Gamma_{3}}{\Gamma_{3}-1} \frac{\boldsymbol{A}_{0} \boldsymbol{\cdot} \boldsymbol{J}_{0}}{\rho_{0} },
\end{eqnarray}
and $\Gamma_{3}$ is the 3rd adiabatic index.
However, the MRI is a weak-field instability, implying magnetic buoyancy/baroclinicity effects are subdominant. For the work presented here, we neglect the contributions of magnetism to entropy (magnetic buoyancy) and consider adiabatic motions; we expect this to be valid for MRI in the NSSL, but studies of MRI in the deep convection zone at high latitudes would need to incorporate these neglected effects.

We choose our particular magnetic field configuration rather than a pure dipole because the radial component $\boldsymbol{e}_{r} \boldsymbol{\cdot} \boldsymbol{B}_{0} = \mathcal{B}(r)\cos(\theta)$ vanishes at $r=r_{1}$. The poloidal field strength in the photosphere is $\approx 1 \, \text{G}$ but measurements suggest sub-surface field strengths $\approx\,$50--150$\,$G\cite{Baldner2009}. The near-surface field should exhibit a strong horizontal (as opposed to radial) character. Magnetic pumping\citesupp{Tobias2001} via surface granulation within the outer 1\% of the solar envelope could account for filtering the outward radial field, with sunspot cores being prominent exceptions. 

We also test pure dipoles and fields with an $\approx 5\%$ dipole contribution, yielding similar results. 
Furthermore, we test that the poloidal field is stable to current-driven instabilities. 
Our chosen confined field also has the advantage that 
$\boldsymbol{e}_{\theta} \boldsymbol{\cdot} \boldsymbol{B}_{0}$ is constant to within $8 \%$ over $r_{0} < r < r_{1}$. However, a pure dipole varies by $\approx 37\%$ across the domain. 
The RMS field amplitude is $|\boldsymbol{B}|_{\text{RMS}} \approx 2\, B_{0} = 180\,\text{G}$, about 25\% larger than the maximum reported inferred dipole equivalent\cite{Baldner2009}. However, projecting our field onto a dipole template gives an $\approx 70\,\text{G}$ equivalent at the $r=r_{1}$ equator. Overall, the subsurface field is the least constrained input to our calculations, the details of which surely change over multiple cycles. 

\textit{Model equations:} Respectively, the linearised anelastic momentum, mass-continuity, and magnetic induction equations are: 
\begin{eqnarray}
\label{momentum}
& \rho_{0} \left( \partial_{t} \boldsymbol{u}   +  \boldsymbol{\omega}_{0} \boldsymbol{\times}  \boldsymbol{u} + \boldsymbol{\omega} \boldsymbol{\times} \boldsymbol{u}_{0} + \boldsymbol{\nabla} \varpi \right)  \ = \ \nonumber \\
& \qquad \nu  \boldsymbol{\nabla \cdot} \left(\rho_{0}\boldsymbol{\sigma} \right) + 
\boldsymbol{j} \boldsymbol{\times} \boldsymbol{B}_{0} +  \boldsymbol{J}_{0} \boldsymbol{\times} \boldsymbol{b} ,\\
\label{mass}
& \boldsymbol{\nabla \cdot} \left( \rho_{0} \boldsymbol{u} \right) = 0,\\
\label{flux}
& \partial_{t} \boldsymbol{b}  - \eta \nabla^{2} \boldsymbol{b} = \boldsymbol{\nabla\times} \left( \boldsymbol{u}_{0} \boldsymbol{\times} \boldsymbol{b} +  \boldsymbol{u} \boldsymbol{\times} \boldsymbol{B}_{0}\right),
\end{eqnarray}
where the traceless strain rate 
\begin{eqnarray}\boldsymbol{\sigma} \ = \ \boldsymbol{\nabla}{\boldsymbol{u}} + (\boldsymbol{\nabla}{\boldsymbol{u}})^{\!\top} - \frac{2}{3} \boldsymbol{\nabla \cdot}{\boldsymbol{u}}\, \mathbf{I}.
\end{eqnarray}

To find eigenstates, $\partial_{t} \to \gamma + i\, \omega$, where $\gamma$ is the real-valued grown rate, and $\omega$ is a real-valued oscillation frequency. 
The induction equation (\ref{flux}) automatically produces MRI solutions satisfying $\boldsymbol{\nabla \cdot}{\boldsymbol{b}} = 0$.

Given the velocity perturbation, $\boldsymbol{u}$, the vorticity $\boldsymbol{\omega} = \boldsymbol{\nabla\times} \boldsymbol{u}$. 
Given the magnetic field (Gauss in cgs units), the current density perturbations $\boldsymbol{j} = \boldsymbol{\nabla\times} \boldsymbol{b}/4\pi$.
At linear order, the Bernoulli function $\varpi = \boldsymbol{u_{0}}\cdot \boldsymbol{u} + h'$, where $h'$ represent enthaply perturbations\cite{Vasil2021}.

The velocity perturbations are impenetrable ($u_{r} = 0$) and stress-free ($\sigma_{r\theta}=\sigma_{r\phi} = 0$) at both boundaries.
For the magnetic field, we enforce perfect conducting conditions at the inner boundary ($b_{r} = \partial_{r} b_{\theta} = \partial_{r} b_{\phi} = 0$).
At the outer boundary, we test three different choices in common usage, as different magnetic boundary conditions have different implications for magnetic helicity fluxes through the domain, and these can affect global dynamo outcomes \citesupp{Kapyla_et_al_2010}. Two choices with zero helicity flux are perfectly conducting and vacuum conditions, and we find only modest differences in the results. We also test a ``vertical field'' or ``open'' boundary (\textit{i.e.}, $\partial_{r} b_r = b_{\theta} = b_{\phi} = 0$) which, though non-physical, explicitly allows a helicity flux.  These open systems also had essentially the same results as the other two for growth rates and properties of eigenfunctions. We conduct most of our experiments using perfectly conducting boundary conditions, which we prefer on the same physical grounds as the background field.

We set constant and kinematic viscous and magnetic diffusivity parameters $\nu = \eta = 10^{-6}$ in units where $\Omega_{0} =  R_{\odot} = 1$. The magnetic Prandtl number $\nu/\eta = \text{Pm}=1$ assumes equal transport of vectors by the turbulent diffusivities. A more detailed analysis of the shear Reynolds numbers yields $\text{Re} = \text{Rm} = U_{0}\, L_{0} / \nu \approx 1500$ where $U_{0} \approx 5000
\,\text{cm}/\text{s}$ is the maximum shear velocity jump across the NSSL and $L_{0} \approx 0.06 R_{\odot}$ is the distance between min/max shear velocity (see ``NSSL energetics and turbulence parameterisation'' below).

We \textit{a posteriori} compute the following scalar-potential decompositions, 
\begin{eqnarray}
\boldsymbol{u} &=& u_{\phi} \, \boldsymbol{e}_{\phi} + \frac{1}{\rho_{0}} \boldsymbol{\nabla\times} ( \rho_{0}\, \psi\, \boldsymbol{e}_{\phi}). \\ 
\boldsymbol{b} &=& b_{\phi} \, \boldsymbol{e}_{\phi} +  \boldsymbol{\nabla\times} (  a_{\phi} \, \boldsymbol{e}_{\phi}),
\end{eqnarray}
where both the magnetic scalar potential, $a_{\phi}$, and the streamfunction, $\psi$, vanish at both boundaries. 

We, furthermore, compute the current helicity correlation relative to global RMS values, 
\begin{eqnarray}
\mathcal{H} = \frac{\boldsymbol{b}\boldsymbol{\cdot} \boldsymbol{j}}{|\boldsymbol{b}|_{\text{RMS}} \, |\boldsymbol{j}|_{\text{RMS}}}.
\end{eqnarray}

There is no initial helicity in the background poloidal magnetic field, 
\begin{eqnarray}
\boldsymbol{B}_{0} = \boldsymbol{\nabla\times} (A_{0}(r,\theta)\, \boldsymbol{e}_{\phi}) \qquad \implies \qquad  \boldsymbol{B}_{0}\cdot (\boldsymbol{\nabla\times}  \boldsymbol{B}_{0}) \ = \ 0. \nonumber 
\end{eqnarray}
Linear dynamical perturbations, $\boldsymbol{b}(r,\theta)$, will \textit{locally} align with the background field and current.
However, because the eigenmodes are wave-like, these contributions vanish exactly when averaged over hemispheres. 
\begin{eqnarray}
\left<\, \boldsymbol{b} \cdot (\boldsymbol{\nabla\times} \boldsymbol{B}_{0}) \,\right> \ = \  \left<\, \boldsymbol{B}_{0}\cdot (\boldsymbol{\nabla\times}  \boldsymbol{b}) \, \right> \ = \ 0.  \nonumber 
\end{eqnarray}
The only possible hemispheric contributions arise when considering quadratic mode interactions, 
\begin{eqnarray}
\left<\, \boldsymbol{b} \cdot (\boldsymbol{\nabla\times} \boldsymbol{b} ) \, \right>  \ \ne \ 0. \nonumber 
\end{eqnarray}
This order is the first where we could expect a non-trivial signal. 

Finally, we also solve the system using multiple different mathematically equivalent equation formulations (e.g., using a magnetic vector potential $\boldsymbol{b} = \boldsymbol{\nabla\times} \boldsymbol{a}$, or dividing the momentum equations by $\rho_{0}$). In all cases, we find excellent agreement in the converged solutions. We prefer this formulation because of satisfactory numerical conditioning as parameters become more extreme.

\textit{Computational considerations:} The Dedalus code\cite{Burns2020} employs general tensor calculus in spherical-polar coordinates using spin-weighted spherical harmonics in $(\theta,\phi)$\citesupp{Vasil2019,Lecoanet2019}.  For the finite radial shell, the code uses a weighted generalised Chebyshev series with sparse representations for differentiation, radial geometric factors and non-constant coefficients (e.g., $\rho_{0}(r)$). Since the background magnetic field and the differential rotation are axisymmetric and they only contain a few low-order separable terms in latitude and radius, these two-dimensional non-constant coefficients have a low-order representation in a joint expansion of spin-vector harmonics and Chebyshev polynomials. The result is a two-dimensional generalised non-Hermetian eigenvalue problem $A \, x= \lambda\, B \, x$, where $x$ represents the full system spectral-space state vector. The matrices, $A$ and $B$, are spectral-coefficient representations of the relevant linear differential and multiplication operators. Cases (i) \& (ii) use 384 latitudinal and 64 radial modes (equivalently spatial points). The matrices $A$ and $B$ remain sparse, with respective fill factors of about $8 \!\cdot\! 10^{-4}$ and $4 \!\cdot\! 10^{-5}$. 

The eigenvalues and eigenmodes presented here are converged to better than 1\% relative absolute error (comparing 256 versus 384 latitudinal modes). 
We also use two simple heuristics for rejecting poorly converged solutions. First, because $\lambda_{0}$ is complex-valued, the resulting iterated solutions do not automatically respect Hermitian-conjugate symmetry, which we often find violated for spurious solutions. Second, the overall physical system is reflection symmetric about the equator, implying the solutions fall into symmetric and anti-symmetric classes. Preserving the desired parity is a useful diagnostic tool for rejecting solutions with mixtures of the two parities, which we check individually for each field quantity.
The precise parameters and detailed implementation scripts are available at \url{https://github.com/geoffvasil/nssl_mri}.

\subsection*{Analytic/semi-analytic estimates}

\textit{Local equatorial calculation:} Our preliminary estimates of the maximum poloidal field strength involves solving a simplified equatorial model of the full perturbation equations, setting the diffusion parameters $\nu, \eta \to 0$. Using a Lagrangian displacement vector, $\boldsymbol{\xi}$, in Eulerian coordinates
\begin{eqnarray}
\boldsymbol{u} &=& \partial_{t} \boldsymbol{\xi} + \boldsymbol{u}_{0} \boldsymbol{\cdot} \boldsymbol{\nabla} \boldsymbol{\xi} - \boldsymbol{\xi}  \boldsymbol{\cdot} \boldsymbol{\nabla} \boldsymbol{u}_{0} \\ 
\boldsymbol{b} &=& \boldsymbol{\nabla\times} \left(\boldsymbol{\xi} \boldsymbol{\times} \boldsymbol{B}_{0} \right).
\end{eqnarray}
In local cylindrical coordinates near the equator $(r,\phi,z)$, we assume all perturbations are axis-symmetric and depend harmonically $\sim e^{i (k_{z} z - \omega t)}$. The cylindrical assumption simplifies the analytical calculations while allowing a transference of relevant quantities from the more comprehensive spherical model. 
That is, we assume a purely poloidal background field with the same radial form as the full spherical computations, $\boldsymbol{B}_{0} = B_{z}(r) \boldsymbol{e}_{z}$. We use the same radial density and angular rotation profiles, ignoring latitudinal dependence. The radial displacement, $\xi_{r}$, determines all other dynamical quantities,
\begin{eqnarray}
\xi_{\phi} &=& -\frac{2 i \omega  \, \Omega }{\omega^{2}-k_{z}^{2} v_{A}^{2}} \xi_{r}, \\ 
\xi_{z} &=&  \frac{i}{k_{z} \, r\,\rho_{0}} \frac{d(r \rho_{0}  \xi_{r})}{dr}
 \\ 
\varpi &=& v_{A}^{2} \frac{B_{z}'}{B_{z}} \xi_{r} +  \frac{\omega^{2}}{k_{z}^{2} \, r\,\rho_{0}} \frac{d(r \rho_{0}  \xi_{r})}{dr},
\end{eqnarray}
where $v_{A}(r) = B_{z}(r)/\sqrt{4\pi\rho_{0}(r)}$.
The radial momentum equation gives a 2nd-order two-point boundary-value problem for $\xi_{r}(r)$. The resulting real-valued differential equation depends on $\omega^{2}$; the instability transitions directly from oscillations to exponential growth when $\omega=0$.
We eliminate terms containing $\xi_{r}'(r)$ with the Liouville transformation $\Psi(r) = \sqrt{r} B_{z}(r) \xi_{r}(r)$. The system for the critical magnetic field reduces to a Schr\"{o}dinger-type equation,
\begin{eqnarray}
-\Psi''(r) + k_{z}^{2}\, \Psi(r) + V(r)\,  \Psi(r) \ = \ 0, \label{Schrdinger}
\end{eqnarray}
with boundary conditions 
\begin{eqnarray}
\Psi(r=r_{0}) \ = \  \Psi(r=r_{1}) \ = \  0
\end{eqnarray}
and potential,
\begin{eqnarray}
V = \frac{r}{v_{A}^{2}} \frac{d \Omega^{2}}{dr} + \frac{r \rho_{0}} {B_{z}} \frac{d}{dr}\!\!\left(\!\frac{1}{r \rho_{0}} \frac{d B_{z}}{dr} \!\right) + \frac{3}{4 r^{2}}. \label{ideal-eigen}
\end{eqnarray}
\textit{Upper-bound:} The maximum background field strength occurs in the limit $k_{z}\to 0$.
With fixed functional forms for $\Omega(r)$, $\rho_{0}(r)$, we suppose 
\begin{eqnarray}
B_{z}(r) \ = \  B_{1}\, \frac{1+4 (r/r_{1})^5}{5 (r/r_{1})^3},
\end{eqnarray}
with $B_{1}=B_{z}(r_{1})$ setting and overall amplitude and $1/B_{1}^{2}$ serving as a generalised eigenvalue parameter.  We solve the resulting system with Dedalus using both Chebyshev and Legendre series for 64, 128, and 256 spectral modes, all yielding the same result, $B_{1} = 1070\,\text{G}$.  The results are also insensitive to detailed changes in the background profile's functional form. 

\textit{Growth rate:} We use a simplified formula for the MRI exponential growth, $\sim e^{\gamma t}$, in a regime not extremely far above onset\cite{Vasil2015}. That is,
\begin{eqnarray}
\gamma^{2} \ \approx \ \frac{\alpha^{2} \omega_{A}^{2} \, ( 2  \Omega S - \omega_{A}^{2}  \,(1+\alpha^{2})) }{\omega_{A}^{2} + 4\Omega^{2}},
\end{eqnarray}
where $\alpha = 2\,H/L \approx 0.2\text{--}0.3$ is the mode aspect ratio with latitudinal wavelength, $L \approx 20^{\circ}\text{--}30^{\circ} R_{\odot}$, and NSSL depth $H \approx 0.05 \,R_{\odot}$. The main text defines all other parameters. In the NSSL $S \approx \Omega$. Therefore $\gamma/\Omega \approx 0.1$ when $\alpha \approx 0.3$, $\omega_{A}/\Omega \approx 1$, and $\gamma/\Omega \approx 0.01$ when $\alpha \approx 0.2$, $\omega_{A}/\Omega \approx 0.1$.

\textit{Saturation amplitude}: We use non-dissipative quasi-linear theory\cite{Vasil2015} to estimate the amplitude of the overall saturation. In a finite-thickness domain, the MRI saturates by transporting mean magnetic flux and angular momentum radially. Both quantities are (approximately) globally conserved; however, the instability shifts magnetic flux inward and angular momentum outward, so the potential from eq.~(\ref{ideal-eigen}) is positive everywhere in the domain. 

Given the \textit{cylindrical} radius, $r$, the local angular momentum and magnetic flux density 
\begin{eqnarray}
L = \rho_{0}\, r\, u_{\phi} , \qquad M = \rho_{0}\, r\, a_{\phi}.
\end{eqnarray}
The respective local flux transport 
\begin{eqnarray}
\partial_{t} L + \boldsymbol{\nabla \cdot}(L \boldsymbol{u}) &=& \boldsymbol{\nabla \cdot}(r\,  b_{\phi} \boldsymbol{b}), \\
\partial_{t} M + \boldsymbol{\nabla \cdot}(M \boldsymbol{u}) &=& 0. 
\end{eqnarray}
For quadratic-order feedback,
\begin{eqnarray}
    \partial_{t}  (\rho_{0} r^{2} \delta u_{\phi})  &=& \partial_{r} ( r^{2} ( b_{\phi} b_{r} -   \rho_{0} u_{\phi} u_{r}  )) +  \partial_{z} ( r^{2} ( b_{\phi} b_{z} -   \rho_{0} u_{\phi} u_{z}  )), \\ 
    \partial_{t}  (\rho_{0} r^{2} \delta a_{\phi})  &=& -\partial_{r} ( r^{2} \rho_{0} a_{\phi} u_{r}  ))  -\partial_{z} ( r^{2} \rho_{0} a_{\phi} u_{z}  )).
\end{eqnarray}
For linear meridional perturbations, 
\begin{eqnarray}
    u_{r} \ = \  -\partial_{z} \psi, & &  
     u_{z} \ = \   \frac{\partial_{r}(r \rho_{0} \,\psi)}{r \rho_{0}},
\\
    b_{r} \ = \  -\partial_{z} a_{\phi}, & &  
     b_{z} \ = \   \frac{\partial_{r}(r  \,a_{\phi})}{r}.
\end{eqnarray}
For the angular components,
\begin{eqnarray}
    \partial_{t}  u_{\phi}  &=& \partial_{z} \! \left( \! (2 \Omega - S) \, \psi +  \frac{B_{z}}{4\pi \rho_{0}} b_{\phi} \right), \\ 
    \partial_{t}  a_{\phi} &=& \partial_{z}( B_{z} \psi ),\\ 
    \partial_{t}b_{\phi}  & = &  \partial_{z} \left(B_{z} \,u_{\phi}+S \, a_{\phi} \right).
\end{eqnarray}
Using the linear balances, we time integrate to obtain the latitudinal-mean rotational and magnetic feedback,
\begin{eqnarray}
    \delta\Omega &=&  \frac{1}{r^{3} \rho_{0}} \partial_{r} \left( r^{2} \rho_{0}\, \mathcal{L} \right), \\ 
    \delta A &=& \frac{1}{r^{2} \rho_{0}}\partial_{r}\left( r^{2} \rho_{0}  \, \Phi \right).
\end{eqnarray}
where angle brackets represent $z$ averages and 
\begin{eqnarray}
    \mathcal{L} &=& \frac{2 B_{z} \langle a_{\phi} u_{\phi} \rangle - (2\Omega-S) \,  \langle a_{\phi}^{2} \rangle }{2 B_{z}^{2}}, \\ 
    \Phi &=&  \frac{\langle a_{\phi}^{2}   \rangle}{2B_{z}}.
\end{eqnarray} 
The dynamic shear and magnetic corrections,
\begin{eqnarray}
    \delta S \ = \ - r\, \partial_{r} \delta \Omega, \qquad \delta B_{z} \ = \ \frac{1}{r}\partial_{r}(r \delta A).
\end{eqnarray}

We derive an overall amplitude estimate by considering the functional
\begin{eqnarray}
    \mathcal{F} = 
    \int ( V\, |\Psi|^{2} + |\boldsymbol{\nabla} \Psi|^{2}) \, \mathrm{d} r,
\end{eqnarray}
which results from integrating eq.~(\ref{Schrdinger}) with respect to $\Psi^{*}(r)$. The saturation condition is 
\begin{eqnarray}
\delta \mathcal{F} = - \mathcal{F}.
\end{eqnarray}
The left-hand side includes all linear-order perturbations in the potential, $\delta V$, and wavefunction, $\delta \Psi$, where 
\begin{eqnarray}
\delta V = \frac{2r}{v_{A}^{2}} \frac{d (\Omega \delta \Omega) }{dr} - 2\frac{\delta B_{z} } {B_{z}} \frac{r}{v_{A}^{2}} \frac{d \Omega^{2} }{dr}      + \frac{r \rho_{0}} {B_{z}} \frac{d}{dr}\!\!\left(\!\frac{1}{r \rho_{0}} \frac{d \delta B_{z}}{dr} \!\right) - \frac{\delta B_{z}}{B_{z}} \frac{r \rho_{0}} {B_{z}} \frac{d}{dr}\!\!\left(\!\frac{1}{r \rho_{0}} \frac{d  B_{z}}{dr} \!\right),
\end{eqnarray}
\begin{eqnarray}
\delta \Psi = \frac{\delta B_{z}}{B_{z}} \, \Psi.
\end{eqnarray}
All reference and perturbation quantities derive from the full sphere numerical eigenmode calculation. We translate to cylindrical coordinates by approximating $z$-averages with latitudinal-$\theta$ averages. The spherical eigenmodes localise near the equator, and the NSSL thickness is only $\approx 5\%$ of the solar radius, justifying the cylindrical approximation in the amplitude estimate. 

Empirically, the first $\delta V$ term dominates the overall feedback calculation, owing to the shear corrections $\propto d \delta \Omega/dr \sim 1/H_{r}^{2}$. Isolating the shear effect produces the simple phenomenological formula in eq.~(\ref{saturation}).

\subsection*{NSSL energetics and turbulence parameterisation}

We estimate that the order-of-magnitude energetics of the NSSL are consistent with the amplitudes of torsional oscillations.
The torsional oscillations comprise $|\Omega'| \approx 1\,$nHz rotational perturbation, relative to the $\Omega_{0} \approx 466$ nHz equatorial frame rotation rate. However, the NSSL contains $\Delta \Omega \approx 11\,$nHz mean rotational shear estimated from the functional form in eqs.~(\ref{eq: omega_fit start})-(\ref{eq: omega_fit end}).  In terms of velocities, the shear in the NSSL has a peak contrast of roughly $U_{0} \approx 5\cdot 10^{3}\,$cm/s across a length scale $L_{0} \approx 0.06 R_\odot$. The relative amplitudes of the torsional oscillations to the NSSL background, $|\Omega'|/\Delta \Omega$, are thus about 10\%. Meanwhile, the radial and latitudinal global differential rotation have amplitudes of order $\approx 100\,$nHz. The relative energies are approximately the squares of these, implying that the $\Delta \textit{KE}$ of the torsional oscillations is $\sim 0.01\%$ to the differential rotation and $\sim 1\%$  to the NSSL. These rough estimates show that the NSSL and the differential rotation can provide ample energy reservoirs for driving an MRI dynamo, and the amplitude of the torsional oscillations is consistent with nonlinear responses seen in classical convection zone dynamos\cite{Nelson2011}.

Vigorous hydrodynamic convective turbulence likely establishes the differential rotation of the NSSL. The large reservoir of shear in the solar interior plays the analogue role of gravity and Keplarian shear in accretion disks. The details of solar convection are not well understood nor well-constrained by observations.  
There are indications, however, that the maintenance of the NSSL is separate from the solar cycle since neither the global differential rotation nor the NSSL shows substantial changes during the solar cycle other than in the torsional oscillations. 

Strong dynamical turbulence in the Sun's outer layers is an uncertainty of our MRI dynamo framework, but scale separation gives hope for progress. From our linear instability calculations, the solar MRI operates relatively close to onset and happens predominantly on large scales. If the fast turbulence of the Sun's outer layers acts mainly as an enhanced dissipation, then the solar MRI should survive relatively unaffected. Treating scale-separated dynamics in this fashion has good precedent: large-scale baroclinic instability in the Earth's atmosphere gracefully ploughs through the vigorous moist tropospheric convection (e.g., thunderstorms). Scale-separated dynamics are particularly relevant because the MRI represents a type of \textit{essentially nonlinear dynamo}, which cyclically reconfigures an existing magnetic field with kinetic energy as a catalyst. From prior work, it is clear that the deep solar convection zone can produce global-scale fields, but those generally have properties very different from the observed fields\cite{Nelson2011}. Essentially nonlinear MHD dynamos have analogues in pipe turbulence. Like those systems, the self-sustaining process leads to an attractor where the dynamo settles into a cyclic state independent of its beginnings.

A full nonlinear treatment of turbulence in the NSSL-MRI setting awaits future work. Here, we adopt a simplified turbulence model via enhanced dissipation. To model the effects of turbulence, we assume that the viscous and magnetic diffusivities are enhanced such that the turbulent magnetic Prandtl number $\text{Pm}=1$ (with no principle of turbulence suggesting otherwise). The momentum and magnetic Reynolds numbers are $\text{Re}=\text{Rm} \approx 1.5 \cdot 10^{3}$. These values are vastly more dissipative than the microphysical properties of solar plasma ($\text{Re} \sim 10^{12}$), and the microphysical $\text{Pm} \ll 1$, implying that $\text{Rm} \ll \text{Re}$.  The studies conducted here find relative independence in the MRI on the choices of $\text{Re}$ within a modest range. In contrast, other instabilities (e.g., convection) depend strongly on $\text{Re}$. We compute sample simulations down to $\text{Re} \approx 50$ with qualitatively similar results, though they match the observed patterns less well and require somewhat stronger background fields. Our adopted value of $\text{Re}\approx 1500$ strikes a good balance for an extremely under-constrained process. Our turbulent parameterisations also produce falsifiable predictions: our proposed MRI dynamo mechanism would face severe challenges if future helioseismic studies of the Sun suggest that the turbulent dissipation is much larger than expected (e.g. if the effective $\text{Re} \ll 1$).  But it is difficult to imagine how any nonlinear dynamics would happen in this scenario.

\def\refname{Additional References for Methods}

\pagebreak

\textbf{Data availability:} Raw eigenfunction/eigenvalue data used to generate Figure 2 can be found along with the analysis scripts at: \url{https://github.com/geoffvasil/nssl_mri}. 

\textbf{Code availability:} We use the Dedalus code and additional analysis tools written in Python, as noted and referenced in Methods. Beyond the main Dedalus installation, all scripts are available at: \url{https://github.com/geoffvasil/nssl_mri}.

\textbf{Acknowledgements}: D.L., K.A., K.J.B., J.S.O., and B.P.B. are supported in part by NASA HTMS grant 80NSSC20K1280. D.L., K.J.B., J.S.O., and B.P.B. are partly supported by NASA OSTFL grant 80NSSC22K1738. N.B.\ acknowledges support from NASA grant 80NSSC22M0162. Computations were conducted with support by the NASA High-End Computing Program through the NASA Advanced Supercomputing (NAS) Division at Ames Research Center on the Pleiades supercomputer with allocation GID s2276.

\textbf{Author Contributions:} 
G.M.V., D.L., J.S.O., and K.J.\ conceived and designed the experiments. G.M.V.\ performed analytical and semi-analytical estimates. K.J.B.\ and D.L.\ developed the multidimensional spherical eigenvalue solver used in the computations. K.A.\ and D.L.\ performed and verified the eigenvalue computations. K.J.B., G.M.V., J.S.O., D.L.\ and B.P.B.\ contributed to the numerical simulation and analysis tools. G.M.V., D.L., B.P.B. and J.S.O.\ wrote and, along with K.J.B., K.A., K.J.\ and N.B., revised the manuscript. All authors participated in discussions around the experimental methods, results and the manuscript. 

\textbf{Competing Interests:} The authors declare no competing interests.

\textbf{Additional Information:} Correspondence and requests for materials should be addressed to G.M.V. Reprints and permissions information are available at www.nature.com/reprints.

\pagebreak

\section*{Extended Data}

\vspace{1cm}

\begin{figure}[H]
\begin{center}
\begin{tabular}{c} \\
\includegraphics[width=15cm]{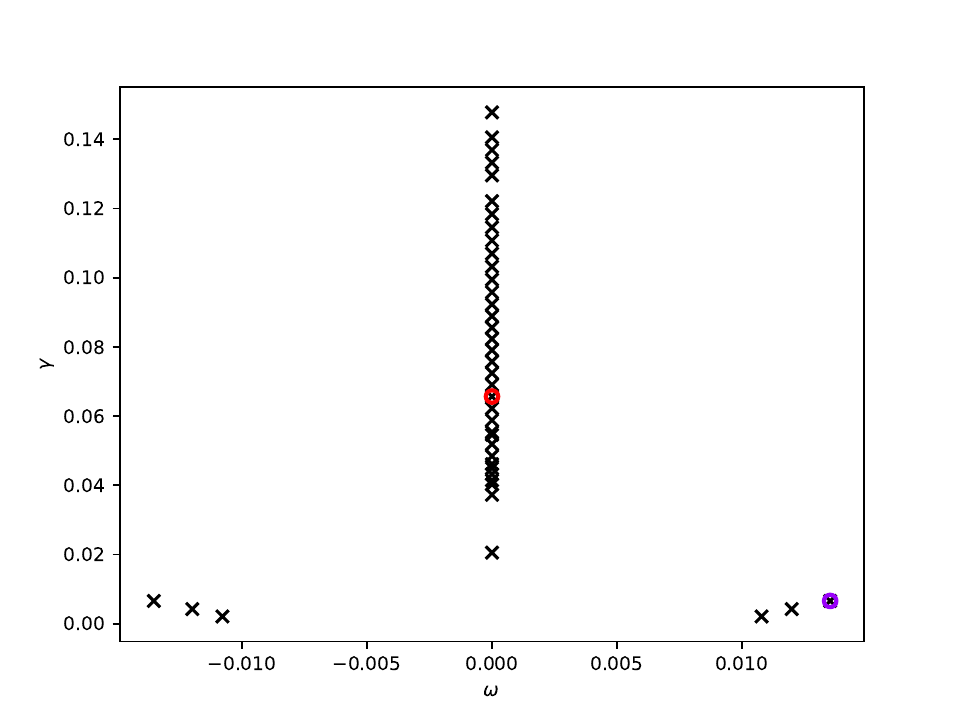}
\end{tabular}
\end{center}
\vspace{0.0cm}
\singlespace 
\caption{Full diagram of complex-valued eigen-spectrum. The time dependence sends $\partial_{t} \to  \gamma + i \omega$, with each real/imaginary component measured in terms of $\Omega_{0} = 466 \, \text{nHz}$. The modes along the vertical axis appear to lie on a continuum, accumulating at a lower value. The isolate modes appear to be point spectra. The red circle represents the case (i) ``fast branch'' from the main text. The purple circle (with its complex conjugate) represent the case (ii) presented ``slow branch''.}
\end{figure}

\pagebreak

\begin{figure}[H]
\begin{center}
\begin{tabular}{c} \\
\includegraphics[width=15cm]{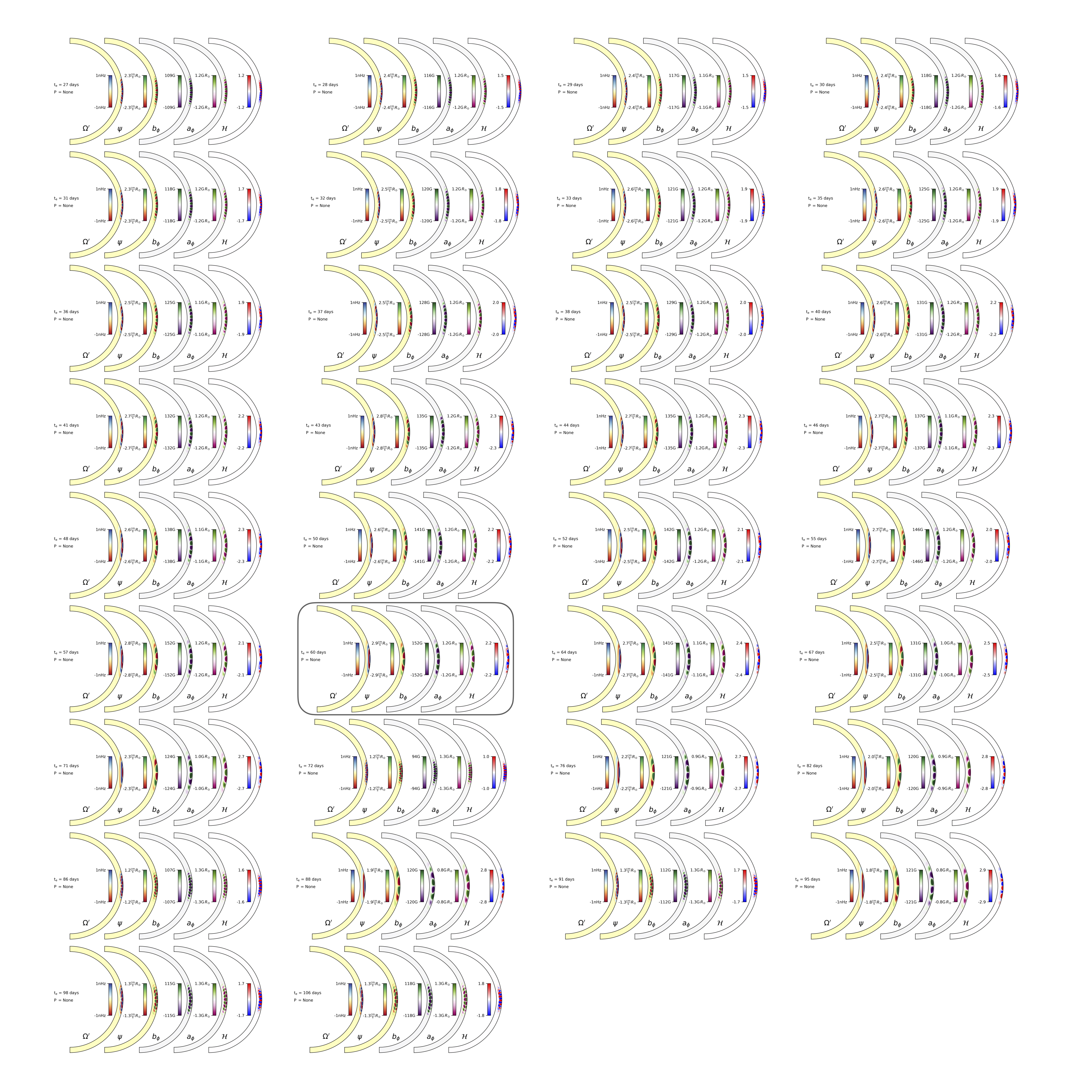}
\end{tabular}
\end{center}
\vspace{0.0cm}
\singlespace 
\caption{The complete collection of ``fast branch'' modes. The growth rates correspond to the vertical axis of Extended Data fig.\,1. Each case contains no discernible oscillations. For completeness, we show (boxed in grey) the $t_{e} = 60\,\text{day}$ fast-branch case (i) presented in the main text fig.~2(a) }
\end{figure}

\pagebreak

\begin{figure}
\begin{center}
\begin{tabular}{c} \\
\includegraphics[width=15cm]{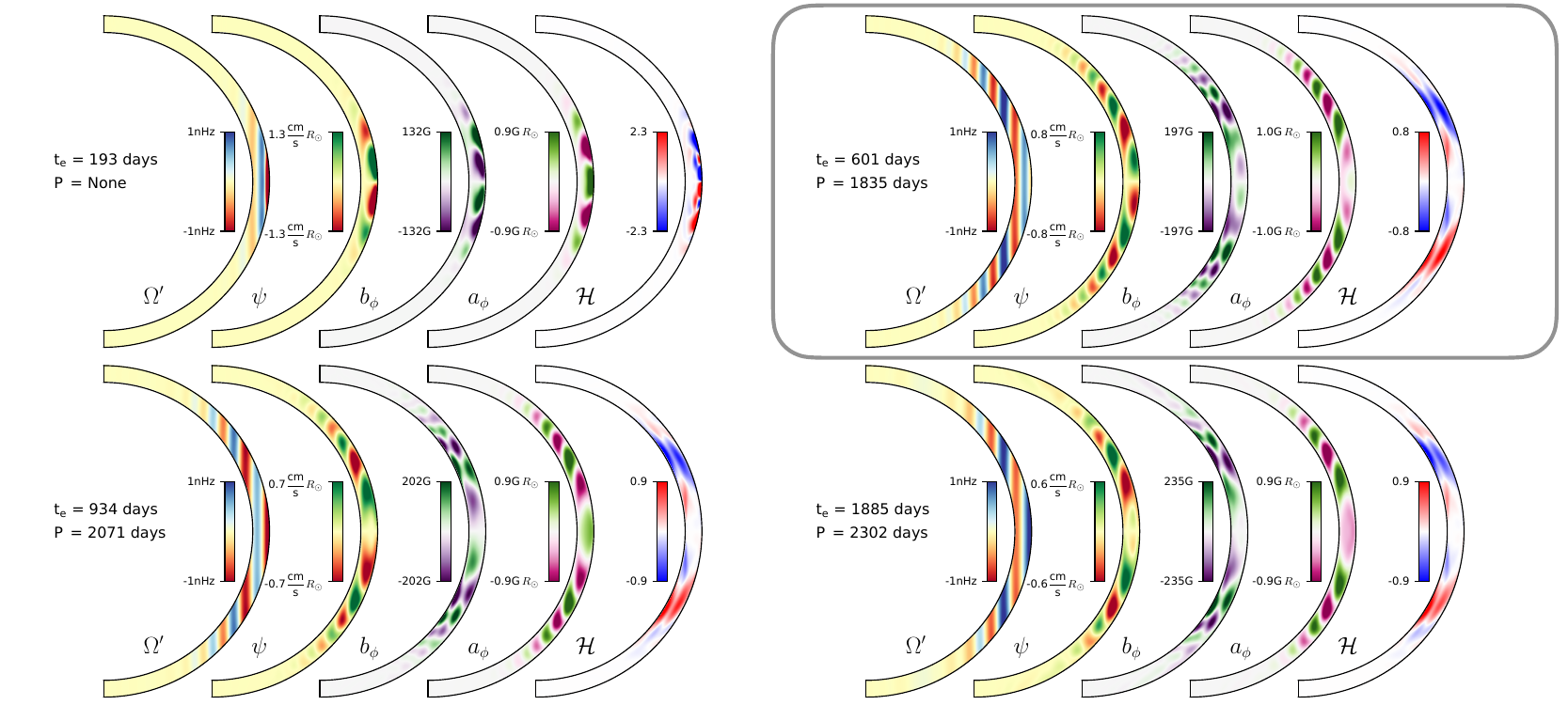}
\end{tabular}
\end{center}
\vspace{0.0cm}
\singlespace 
\caption{The complete collection of ``slow branch'' modes. The growth rates correspond to the isolated spectrum in Extended Data fig.\,1. The upper-left image shows the point spectra along the vertical axis. The three other images show the isolated oscillatory modes, including the slow-branch case (ii) mode (boxed in grey) presented in the main text fig.~2(b).}
\end{figure}

\end{document}